\documentclass[12pt, onecolumn, draftclsnofoot]{IEEEtran}

\usepackage{amssymb, amsmath}
\usepackage[noadjust]{cite}
\usepackage{url}

\newtheorem{theorem}{Theorem}
\newtheorem{lemma}[theorem]{Lemma}
\newtheorem{corollary}[theorem]{Corollary}
\newtheorem{definition}[theorem]{Definition}
\newtheorem{example}[theorem]{Example}
\newtheorem{construction}[theorem]{Construction}

\newcommand{\beqn}{\begin{equation}}
\newcommand{\eeqn}{\end{equation}}
\newcommand{\beq}{\begin{equation*}}
\newcommand{\eeq}{\end{equation*}}

\newcommand{\Z}{\mathbb Z}
\newcommand{\Cn}{\mathbb C}
\newcommand{\C}{{\cal C}}
\newcommand{\A}{{\cal A}}
\newcommand{\R}{{\cal R}}
\newcommand{\F}{{\cal F}}

\newcommand{\RM}{{\rm RM}}
\newcommand{\ZRM}{{\rm ZRM}}
\newcommand{\ERM}{{\rm ERM}}
\newcommand{\PMEPR}{\mbox{PMEPR}}

\newcommand{\bia}{{\boldsymbol{a}}}
\newcommand{\bib}{{\boldsymbol{b}}}
\newcommand{\bix}{{\boldsymbol{x}}}
\newcommand{\bic}{{\boldsymbol{c}}}
\newcommand{\bid}{{\boldsymbol{d}}}

\newcommand{\biA}{{\boldsymbol{A}}}
\newcommand{\biB}{{\boldsymbol{B}}}
\newcommand{\biC}{{\boldsymbol{C}}}
\newcommand{\biF}{{\boldsymbol{F}}}

\newcommand{\bnull}{{\boldsymbol{0}}}

\newcommand{\wt}{{\rm wt}}
\newcommand{\dist}{{\rm d}}

\begin{document}

\title{Complementary Sets, Generalized Reed--Muller Codes, and Power Control for OFDM}

\author{Kai-Uwe Schmidt\thanks{The author is with Communications Laboratory, Dresden University of Technology, 01062 Dresden, Germany, e-mail: schmidtk@ifn.et.tu-dresden.de, web: {http://www.ifn.et.tu-dresden.de/$\sim$schmidtk/}}}

\maketitle

\begin{abstract}
The use of error-correcting codes for tight control of the peak-to-mean envelope power ratio (PMEPR) in orthogonal frequency-division multiplexing (OFDM) transmission is considered in this correspondence. By generalizing a result by Paterson, it is shown that each $q$-phase ($q$ is even) sequence of length $2^m$ lies in a complementary set of size $2^{k+1}$, where $k$ is a nonnegative integer that can be easily determined from the generalized Boolean function associated with the sequence. For small $k$ this result provides a reasonably tight bound for the PMEPR of $q$-phase sequences of length $2^m$. A new $2^h$-ary generalization of the classical Reed--Muller code is then used together with the result on complementary sets to derive flexible OFDM coding schemes with low PMEPR. These codes include the codes developed by Davis and Jedwab as a special case. In certain situations the codes in the present correspondence are similar to Paterson's code constructions and often outperform them.
\end{abstract}



\section{Introduction}

In some applications the advantages of the orthogonal frequency-division multiplexing (OFDM) modulation technique are outweighed by the typically high peak-to-mean envelope power ratio (PMEPR) of uncoded OFDM signals. Among various approaches to solve this power-control issue, the use of block coding across the subcarriers \cite{Jones1994}, \cite{Jones1996} is one of the more promising concepts \cite{Paterson2000}. Here the goal is to design error-correcting codes that contain only codewords with low PMEPR.
\par
Sequences lying in \emph{complementary pairs} \cite{Golay1961}, also called \emph{Golay sequences}, are known to have PMEPR at most $2$ in $q$-ary phase-shift keying (PSK) modulation \cite{Popovic1991}. In \cite{Davis1999} Davis and Jedwab developed a powerful theory linking Golay sequences with generalized Reed--Muller codes. More specifically, it was shown that a family of binary Golay sequences of length $2^m$ organizes in $m!/2$ cosets of $\RM_{2}(1,m)$ inside $\RM_{2}(2,m)$, where $\RM_2(r,m)$ is the Reed--Muller code of order $r$ and length $2^m$ \cite{MacWilliams1977}. Similarly for $h>1$ \cite{Davis1999} identifies $m!/2$ cosets of $\RM_{2^h}(1,m)$ comprised of polyphase Golay sequences inside $\ZRM_{2^h}(2,m)$. Here $\RM_{2^h}(r,m)$ and $\ZRM_{2^h}(r,m)$ are generalizations of the classical Reed--Muller code over ${2^h}$-ary alphabets. For small $m$, say $m\le 5$, the union of these cosets yields powerful code with good error-correcting properties and strictly bounded PMEPR.
\par
However the rate of this code rapidly tends to zero when the block length increases. Therefore Davis and Jedwab proposed \cite{Davis1999} to include further cosets of $\RM_{2^h}(1,m)$ in order to increase the code rate at the cost of a slightly larger PMEPR. While in \cite{Davis1999} such cosets have been identified with an exhaustive search, a more sophisticated theory was developed by Paterson in \cite{Paterson2000a}; it was shown that each coset of $\RM_{2^h}(1,m)$ inside $\RM_{2^h}(2,m)$ can be partitioned into \emph{complementary sets} of size $2^{k+1}$, where $k$ is a nonnegative integer that can be easily determined from a representative of the coset. Since the PMEPR of each sequence lying in a complementary set of size $N$ has PMEPR at most $N$, \cite{Paterson2000a} provides an upper bound on the PMEPR of arbitrary second-order cosets of $\RM_{2^h}(1,m)$. This result was then exploited in various ways to obtain coding schemes for OFDM, which extend those proposed in \cite{Davis1999}.
\par
Several further constructions linking complementary sets and Reed--Muller codes have been described in \cite{Stinchcombe2000}, \cite{Parker2003}, \cite{Chen2006}. Although these results often provide better upper bounds on the PMEPR than the work in \cite{Paterson2000a}, it seems difficult to use them to derive practicable coding schemes for OFDM.
\par
In this correspondence we generalize the results from \cite{Paterson2000a}. We will establish a construction of sequences that are contained in higher-order generalized Reed--Muller codes and lie in complementary sets of a given size. It appears that \cite[Theorem~12]{Paterson2000a} is a special case of this result. We will then relate this construction to a new generalization of the classical Reed--Muller code, which we call the \emph{effective-degree Reed--Muller code}. In this way, we derive a number of new flexible OFDM coding schemes with low PMEPR. In contrast to the work in \cite{Paterson2000a}, all these codes arise in a uniform way from a general framework. Moreover they often outperform the coding options presented in \cite{Paterson2000a}. The proposed codes are unions of cosets of a linear code over $\Z_{2^h}$ that contains in general more codewords than $\RM_{2^h}(1,m)$, although this linear code itself is a union of cosets of $\RM_{2^h}(1,m)$. Compared to the approaches in \cite{Davis1999} and \cite{Paterson2000a}, this makes our codes more amenable to efficient encoding and decoding algorithms.
\par
The remainder of this correspondence is organized as follows. In the next section we describe a simplified OFDM model, establish our main notation (which essentially follows that in \cite{Paterson2000a}), and present some known results from \cite{Paterson2000a}. Section~\ref{sec:sets} contains our results on complementary sets. In Section~\ref{sec:codes} we introduce the effective-degree Reed--Muller code and derive OFDM codes with low PMEPR. We close with a discussion in Section~\ref{sec:discussion}.


\section{Preliminaries}

\subsection{The OFDM Coding Problem}

We consider an OFDM system with $n$ subcarriers. The transmitted OFDM signal corresponding to the codeword $\biC=(C_0,C_1,\dots,C_{n-1})\in\Cn^n$ is the real part of the \emph{complex envelope}, which can be written as
\beq
S(\biC)(\theta)=\sum\limits_{i=0}^{n-1}C_i\,e^{\,\sqrt{-1} 2\pi (i+\zeta) \theta},\quad 0\le \theta< 1,
\eeq
where $\zeta$ is a positive constant. In the following it is assumed that the elements of $\biC$ are taken from a $q$-ary PSK constellation, i.e., $C_i=\xi^{c_i}$ with $\xi=e^{\sqrt{-1}2\pi/q}$ and $c_i\in\Z_q$. Then $\biC$ is a polyphase sequence. This assumption together with Parseval's identity implies that the complex envelope has mean power equal to $n$. The PMEPR of the codeword $\biC$ (or of the corresponding complex envelope) is then defined to be
\beq
\PMEPR(\biC)\triangleq\frac{1}{n}\sup_{0\le \theta<1} |S(\biC)(\theta)|^2.
\eeq
The PMEPR is always less than or equal to $n$, where the maximum occurs, for example, if $\biC$ is the all-one word. We aim at constructing codes $\C$ that have error-correcting capabilities and for which the value
\beq
\max_{\biC\in\C}\PMEPR(\biC)
\eeq
is substantially lower than $n$.

\subsection{Aperiodic Correlations and Complementary Sets}

Given two complex-valued sequences $\biA=(A_0,A_1\dots,A_{n-1})$ and $\biB=(B_0,B_1\dots,B_{n-1})$ of length $n$, their {\em aperiodic cross-correlation} at a displacement $\ell\in\Z$ is defined to be
\beq
C(\biA,\biB)(\ell)\triangleq 
\begin{cases}
\sum\limits_{i=0}^{n-\ell-1}A_{i+\ell}B^*_i&0\le \ell<n\\
\sum\limits_{i=0}^{n+\ell-1}A_iB^*_{i-\ell}&-n< \ell<0\\
0&\mbox{otherwise,}
\end{cases}
\eeq
where $(.)^*$ denotes complex conjugation. The {\em aperiodic auto-correlation} of $\biA$ at a displacement $\ell\in\Z$ is defined as 
\beq
A(\biA)(\ell)\triangleq C(\biA,\biA)(\ell).
\eeq
\par
\begin{definition}
\label{def:cs}
A set of $N$ sequences $\{\biA^0,\biA^1,\dots,\biA^{N-1}\}$ is a {\em complementary set of size $N$} if
\beq
\sum\limits_{i=0}^{N-1}A(\biA^i)(\ell)=0\quad \mbox{for each}\;\; \ell\ne 0.
\eeq
If $N=2$, the set is called a {\em complementary pair} (or {\em Golay complementary pair}) \cite{Golay1961} and the sequences therein {\em Golay sequences}.
\end{definition}
\par
Golay sequences have found applications in many different areas of signal processing. The work of Popovi\'{c} \cite{Popovic1991}, where it was essentially proved that polyphase Golay sequences have PMEPR at most $2$, motivated the use of such sequences as codewords in OFDM \cite{Wilkinson1995}, \cite{Nee1996}, \cite{Ochiai1997}, \cite{Davis1999}. Paterson \cite{Paterson2000a} generalized these results by proving:
\begin{theorem}[\cite{Paterson2000a}]
\label{res:PMEPR-set}
Each polyphase sequence lying in a complementary set of size $N$ has PMEPR at most $N$.
\end{theorem}


\subsection{Generalized Boolean Functions, Associated Sequences and their Correlations}

A {\em generalized Boolean function} $f$ is defined as a mapping $f\,:\,\{0,1\}^m\rightarrow\Z_q$. Such a function can be written uniquely in the polynomial form
\beq
f(x_0,x_1,\dots,x_{m-1})=\sum_{i\in\{0,1\}^m}c_i\, \prod_{\alpha=0}^{m-1}x_\alpha^{i_\alpha},\quad c_i\in\Z_q,
\eeq
called the \emph{algebraic normal form} of $f$. Sometimes we write $f$ in place of $f(x_0,x_1,\dots,x_{m-1})$. If $c_i=1$ for exactly one $i$ and zero otherwise, then $f$ is called a \emph{monomial}. Let $\deg(f)$ denote the algebraic degree of $f$.
\par
A generalized Boolean function may be equally represented by sequences of length $2^m$. Therefore suppose $0\le i<2^m$ has binary expansion $(i_0,i_1,\dots, i_{m-1})$ such that $i=\sum_{\alpha=0}^{m-1}i_\alpha2^\alpha$ and $i_\alpha\in\{0,1\}$, and write $f_i=f(i_0,i_1,\dots,i_{m-1})$. We define
\beq
\psi(f)\triangleq (f_0,f_1,\dots,f_{2^m-1})
\eeq
as the {\em $\Z_q$-valued sequence associated with $f$} and
\beq
\Psi(f)\triangleq (\xi^{f_0},\xi^{f_1},\dots,\xi^{f_{2^m-1}})
\eeq 
as the {\em polyphase sequence associated with $f$}, where $\xi=e^{\sqrt{-1}2\pi/q}$.
\par
In what follows we recall the technique of restricting generalized Boolean functions and their associated polyphase sequences. This technique was introduced in \cite{Paterson2000a} in order to expand aperiodic correlations, as we shall see in Lemma~\ref{lem:CCF-Expanding}.
\par
Suppose that $f: \{0,1\}^m\rightarrow\Z_q$ is a generalized Boolean function in the variables $x_0,x_1,\dots,x_{m-1}$, and let $\biF=\Psi(f)$. Let a list of $k$ indices be given by $0\le j_0<j_1<\cdots<j_{k-1}<m$, and write $\bix=(x_{j_0},x_{j_1},\dots,x_{j_{k-1}})$. Let $\bid=(d_0,d_1,\dots,d_{k-1})$ be a binary word of length $k$, and let $(i_0,i_1,\dots,i_{m-1})$ be the binary expansion of $0\le i<2^m$. The {\em restricted sequence} $\biF|_{\bix=\bid}$ is a sequence of length $2^m$ that coincides with $\biF$ at the positions $i$ where 
$i_{j_\alpha}=d_\alpha$ for each $0\le\alpha<k$. Otherwise $\biF|_{\bix=\bid}$ is equal to zero. For $k=0$ we define $\biF|_{\bix=\bid}\triangleq\biF$.
\par
A sequence that is restricted in $k$ variables comprises $2^m-2^{m-k}$ zero entries and $2^{m-k}$ nonzero entries. Those nonzero entries are determined by a function, which is denoted as $f|_{\bix=\bid}$ and called a {\em restricted generalized Boolean function}. This function is a generalized Boolean function in $m-k$ variables and is obtained by replacing the variables $x_{j_\alpha}$ by $d_\alpha$ for all $0\le\alpha<k$ in the algebraic normal form of $f$. The restricted sequence $\biF|_{\bix=\bid}$ is then recovered by associating a polyphase sequence of length $2^{m-k}$ with $f|_{\bix=\bid}$ and inserting $2^m-2^{m-k}$ zeros at the corresponding positions. Similarly to a disjunctive normal form of a Boolean function \cite{MacWilliams1977}, the original function $f$ can be reconstructed from the functions $f|_{\bix=\bid}$ by
\beq
f=\sum_{\bid\in\{0,1\}^k} f|_{\bix=\bid}\prod_{\alpha=0}^{k-1}x_{j_\alpha}^{d_\alpha}(1-x_{j_\alpha})^{(1-d_\alpha)}.
\eeq
\begin{lemma}[\cite{Paterson2000a}]
\label{lem:CCF-Expanding}
Suppose that $f:\{0,1\}^m\rightarrow \Z_q$ is a generalized Boolean function, and let $\biF=\Psi(f)$. Let $0\le j_0<j_1<\cdots<j_{k-1}<m$ be a list of $k$ indices. Write $\bix=(x_{j_0},x_{j_1},\dots,x_{j_{k-1}})$, and let $\bid,\bid_1,\bid_2\in\{0,1\}^k$. Then we have
\beq
A(\biF)(\ell)
=\sum\limits_{\bid}A(\biF|_{\bix=\bid})(\ell)\\
+\sum\limits_{\bid_1\ne\bid_2}C(\biF|_{\bix=\bid_1},\biF|_{\bix=\bid_2})(\ell).
\eeq
\end{lemma}


\subsection{A Known Construction of Complementary Pairs}

Next we recall a construction of complementary pairs from \cite{Paterson2000a}. A quadratic polynomial $f$ over $\Z_q$ in the $\{0,1\}$-valued variables $x_{i_0},x_{i_1},\dots,x_{i_{m-1}}$ is generally given by
\beq
f(x_{i_0},\dots,x_{i_{m-1}})=\sum_{0\le j<k<m}b_{jk}x_{i_j}x_{i_k}+a(x_{i_0},\dots,x_{i_{m-1}}),
\eeq
where $b_{jk}\in\Z_q$ and $a$ is an affine form over $\Z_q$. With each such a polynomial one can associate a labeled graph, denoted by $G(f)$. The vertices of this graph are labeled with $i_0,i_1,\dots,i_{m-1}$, and the edge between vertex $i_j$ and vertex $i_k$ is labeled with $b_{jk}$. 
\par
Such a graph is called a \emph{path} in $m$ vertices if $q$ is even and $m=1$ (then the graph consists of a single vertex) or if $q$ is even, $m\ge 2$, and $f$ is of the form
\beq
f(x_{i_0},\dots,x_{i_{m-1}})=\frac{q}{2}\sum_{\alpha=0}^{m-2}x_{i_{\pi(\alpha)}}x_{i_{\pi(\alpha+1)}}+a(x_{i_0},\dots,x_{i_{m-1}}),
\eeq
where $\pi$ is a permutation of $\{0,1,\dots,m-1\}$. The indices $i_{\pi(0)}$ and $i_{\pi(m-1)}$ are called \emph{end vertices} of the path. If the path consists of a single vertex, this vertex is called an end vertex as well.
\par
We are now in a position to quote:
\begin{theorem}[\cite{Paterson2000a}]
\label{thm:pairs}
Suppose $m>k$. Let $0\le j_0<j_1<\cdots<j_{k-1}<m$ be a list of $k$ indices, write $\bix=(x_{j_0},x_{j_1},\dots,x_{j_{k-1}})$, and let $\bid\in\{0,1\}^k$. Suppose $f: \{0,1\}^m\rightarrow\Z_q$ is a generalized Boolean function such that $f|_{\bix=\bid}$ is quadratic and $G(f|_{\bix=\bid})$ is a path in $m-k$ vertices. Write $\biF=\Psi(f)$ and $\biF'=\Psi(f+(q/2)x_a+c')$. Then $\biF|_{\bix=\bid}$ and $\biF'|_{\bix=\bid}$ form a complementary pair. Here, $a$ is an end vertex of the path $G(f|_{\bix=\bid})$ and $c'\in\Z_q$.
\end{theorem}
\par
In particular, if $k=0$, the preceding theorem identifies  $(m!/2)q^{m+1}$ polyphase sequences lying in complementary pairs \cite[Corollary~11]{Paterson2000a}, which generalizes the original result by Davis and Jedwab \cite[Theorem~3]{Davis1999} from $q$ being a power of $2$ to even $q$.


\section{A Construction of Complementary Sets}
\label{sec:sets}

In what follows we prove that each polyphase sequence of length $2^m$ lies in a complementary set, whose size can be easily determined by inspecting the generalized Boolean function associated with the sequence. 
\par
\begin{theorem}
\label{thm:sets} 
Suppose $m>k$. Let $0\le j_0<j_1<\cdots<j_{k-1}<m$ be a list of $k$ indices, and write $\bix=(x_{j_0},x_{j_1},\dots,x_{j_{k-1}})$. Let $f: \{0,1\}^m\rightarrow\Z_q$ be a generalized Boolean function such that for each $\bid\in\{0,1\}^k$ the restricted function $f|_{\bix=\bid}$ is quadratic and $G(f|_{\bix=\bid})$ is a path in $m-k$ vertices. Then $\Psi(f)$ lies in a complementary set of size $2^{k+1}$, and the PMEPR of $\Psi(f)$ is at most $2^{k+1}$.
\end{theorem}
\begin{proof}
Write $\bid=(d_0,d_1,\dots,d_{k-1})$ and $\bic=(c_0,c_1,\dots,c_{k-1})$. Define
\beq
\biF_{\bic c'}=\Psi\left(f\,+\,\frac{q}{2}\sum\limits_{\alpha=0}^{k-1}c_\alpha\,x_{j_\alpha}+\frac{q}{2}\,c'\,e\right),
\eeq
where $\bic\in\{0,1\}^k$, $c'\in\{0,1\}$,
\beq
e=\sum_{\bid\in\{0,1\}^k}x_{a_\bid}\prod_{\alpha=0}^{k-1}x_{j_\alpha}^{d_\alpha}(1-x_{j_\alpha})^{(1-d_\alpha)},
\eeq
and $a_\bid$ is an end vertex of the path $G(f|_{\bix=\bid})$. We claim that the set
\beq
\left\{\biF_{\bic c'}\,|\,\bic\in\{0,1\}^k,c'\in\{0,1\}\right\},
\eeq
which contains $\Psi(f)$, is a complementary set of size $2^{k+1}$. To prove this, it has to be shown that the sum of auto-correlations $\sum_{\bic,\,c'} A\left(\biF_{\bic c'}\right)(\ell)$ is zero for each $\ell\ne 0$.
We employ Lemma~\ref{lem:CCF-Expanding} and write
\beq
\sum_{\bic,\,c'} A(\biF_{\bic c'})(\ell)=S_1+S_2,
\eeq
where
\begin{align*}
S_1&=\sum\limits_{\bic,\,c'}\sum\limits_{\bid}A(\biF_{\bic c'}|_{\bix=\bid})(\ell)\\
S_2&=\sum\limits_{\bic,\,c'}\sum\limits_{\bid_1\ne\bid_2}C(\biF_{\bic c'}|_{\bix=\bid_1},\biF_{\bic c'}|_{\bix=\bid_2})(\ell).
\end{align*}
We first focus on the term $S_1$, which can be written as
\beq
S_1=\sum\limits_{\bic}\sum\limits_{\bid}\left[A(\biF_{\bic 0}|_{\bix=\bid})(\ell)+A(\biF_{\bic 1}|_{\bix=\bid})(\ell)\right].
\eeq
Note that $e|_{\bix=\bid}=x_{a_\bid}$. Thus the restricted functions corresponding to $\biF_{\bic 0}|_{\bix=\bid}$ and $\biF_{\bic 1}|_{\bix=\bid}$ are of the form
\begin{align*}
&f|_{\bix=\bid}+\frac{q}{2}\sum\limits_{\alpha=0}^{k-1}c_\alpha\,d_\alpha\\
&f|_{\bix=\bid}+\frac{q}{2}\sum\limits_{\alpha=0}^{k-1}c_\alpha\,d_\alpha+\frac{q}{2}\,x_{a_\bid},
\end{align*}
respectively. Notice that the term containing the sum over $\alpha$ is a constant occurring in both functions. Hence, by hypothesis and by Theorem~\ref{thm:pairs}, $\biF_{\bic 0}|_{\bix=\bid}$ and $\biF_{\bic 1}|_{\bix=\bid}$ form a complementary pair. It follows that the inner term of $S_1$ is zero for each $\ell\ne 0$. Thus also $S_1$ itself is zero for each $\ell\ne 0$.
\par
It remains to show that the sum $S_2$ is zero. This part of the proof follows more or less the same reasoning as the second part of the proof of \cite[Theorem~12]{Paterson2000a}.
\end{proof}
\par
We have a number of notes on Theorem~\ref{thm:sets}. If $k=0$, Theorem~\ref{thm:sets} applies to $(m!/2)q^{m+1}$ polyphase sequences lying in complementary pairs. These are exactly those identified by setting $k=0$ in Theorem~\ref{thm:pairs}. For $k>0$ Theorem~\ref{thm:sets} essentially generalizes \cite[Theorem~12]{Paterson2000a}; if $f$ is constrained to be a quadratic generalized Boolean function, then Theorem~\ref{thm:sets} virtually reduces to \cite[Theorem~12]{Paterson2000a}.
\par
The proof of Theorem~\ref{thm:sets} shows that the sequence $\Psi(f)$ lies in a complementary set that can be decomposed into $2^k$ complementary pairs identified by Theorem~\ref{thm:pairs}. By reversing this process, the sequence $\Psi(f)$ can be constructed by interleaving $2^k$ Golay sequences from Theorem~\ref{thm:pairs}. However the sole application of such an interleaving method would not directly admit the construction of sequences corresponding to generalized Boolean functions of a specific degree, which will be required to derive flexible coding schemes in the next section. 
\par
We also remark that it cannot be expected that Theorem~\ref{thm:sets} provides tight PMEPR bounds for each individual sequence, especially when $k$ is large. Indeed \cite{Stinchcombe2000} (in particular \cite[Theorem~3.6]{Stinchcombe2000}) and the recent work \cite{Chen2006} contain significant improvements of Theorem~\ref{thm:sets} in certain situations. However it seems difficult to exploit these results to derive coding schemes that admit efficient encoding and decoding. 
\par
In summary, the usefulness of Theorem~\ref{thm:sets} lies in the fact that it provides a relatively simple method to identify sets of sequences that correspond to generalized Boolean functions of a given (preferably low) degree and whose PMEPR is bounded above by a given power of $2$. 
\par
We close this section with an example for the application of Theorem~\ref{thm:sets}.
\begin{example}
We take $q=2$ and $m=4$. Let $f\,:\,\{0,1\}^4\rightarrow\Z_2$ be given by
\beq
f(x_0,x_1,x_2,x_3)
=x_0x_1x_2+x_0x_1x_3+x_0x_2+x_1x_3+x_2x_3.
\eeq
By restricting $f$ in $x_0$ (i.e., $\bix=(x_0)$), we obtain the two restricted functions
\begin{align*}
f|_{x_0=0}&=x_1x_3+x_2x_3\\
f|_{x_0=1}&=x_1x_2+x_2x_3+x_2,
\end{align*}
which are quadratic and their associated graphs are paths in $3$ vertices. Hence, by Theorem~\ref{thm:sets}, the PMEPR of $\Psi(f)$ is at most 4. By direct computation it can be observed that the true PMEPR of $\Psi(f)$ is approximately $3.32$.
\end{example}


\section{OFDM Codes with Low PMEPR}
\label{sec:codes}

\subsection{The Effective-Degree Reed--Muller Code}

A code of length $n$ over the ring $\Z_{2^h}$ is linear if it is a submodule of $\Z_{2^h}^n$. A coset of a linear code $\C\subseteq\Z_{2^h}^n$ is defined to be $\{\bia+\bic\,|\,\bic\in\C\}$, where $\bia\in\Z_{2^h}^n$ is a representative of this coset. Despite the fact that a linear code $\C$ defined over a ring does not necessarily have a basis, one can associate a generator matrix with $\C$ such that the codewords of $\C$ are all distinct $\Z_{2^h}$-linear combinations of the rows of this matrix. For background on linear codes over rings we refer to \cite{Hammons1994} and \cite{Calderbank1995a}.
\par
In what follows we generalize the classical Reed--Muller codes \cite{MacWilliams1977} to linear codes over $\Z_{2^h}$. We begin with defining the effective degree of a generalized Boolean function.
\begin{definition}
Let $f:\{0,1\}^m\rightarrow\Z_{2^h}$ be a generalized Boolean function. We define the {\em effective degree of $f$} to be
\beq
\max_{0\le i< h}\,\left[\deg\left(f \bmod 2^{i+1}\right)-i\right].
\eeq
\end{definition}
\vspace{1ex}
\par
For instance, the function $f:\{0,1\}^3\rightarrow\Z_8$ given by $f=4x_0x_1x_2+x_1$ has effective degree equal to $1$. Now let $\F(r,m,h)$ be the set of all generalized Boolean functions $\{0,1\}^m\rightarrow\Z_{2^h}$ of effective degree at most $r$. A simple counting argument leads to
\beqn
\label{eqn:num_F}
\log_2\big|\F(r,m,h)\big|=\sum_{i=0}^{r}h{m\choose i}+\sum_{i=1}^{h-1}(h-i){m\choose r+i}.
\eeqn
\begin{definition}
For $0\le r\le m$ we define the \emph{effective-degree Reed--Muller code} as
\beq
\ERM(r,m,h)\triangleq\left\{\psi(f)\,|\,f\in\F(r,m,h)\right\}.
\eeq
\end{definition}
\vspace{1ex}
\par
It follows that $\ERM(r,m,h)$ is a linear code over $\Z_{2^h}$ and, since the effective degree and the algebraic degree coincide for $h=1$, $\ERM(r,m,1)$ is the classical Reed--Muller code \cite{MacWilliams1977}. A generator matrix for $\ERM(r,m,h)$ has rows corresponding to the words associated with monomials in the variables $x_0,x_1,\dots,x_{m-1}$ of degree at most $r$ together with $2^i$ times the monomials of degree $r+i$, where $i=1,\dots,h-1$. For example a generator matrix for $\ERM(0,3,3)$ is given by:
\beq
\begin{bmatrix}
1 & 1 & 1 & 1 & 1 & 1 & 1 & 1\\
0 & 2 & 0 & 2 & 0 & 2 & 0 & 2\\
0 & 0 & 2 & 2 & 0 & 0 & 2 & 2\\
0 & 0 & 0 & 0 & 2 & 2 & 2 & 2\\
0 & 0 & 0 & 4 & 0 & 0 & 0 & 4\\
0 & 0 & 0 & 0 & 0 & 4 & 0 & 4\\
0 & 0 & 0 & 0 & 0 & 0 & 4 & 4
\end{bmatrix}
\begin{matrix}
1\\2x_0\\2x_1\\2x_2\\4x_0x_1\phantom{.}\\4x_0x_2\phantom{.}\\4x_1x_2.
\end{matrix}
\eeq
Now let $\bia=(a_0,a_1,\dots,a_{n-1})$ be a word with elements in $\Z_{2^h}$. The Lee weight of $\bia$ is defined to be
\beq
\wt_L(\bia)\triangleq\sum_{i=0}^{n-1}\min\{a_i,2^h-a_i\},
\eeq
and its squared Euclidean weight (when the entries of $\bia$ are mapped onto a $2^h$-ary PSK constellation) is given by
\beq
\wt^2_E(\bia)\triangleq\sum_{i=0}^{n-1}\left|\xi^{a_i}-1\right|^2,
\eeq
where $\xi=e^{\sqrt{-1}2\pi/2^h}$. Let $\dist_L(\bia,\bib)\triangleq\wt_L(\bia-\bib)$ and $\dist^2_E(\bia,\bib)\triangleq\wt^2_E(\bia-\bib)$ be the Lee and squared Euclidean distance between $\bia,\bib\in\Z_{2^h}^n$, respectively. Here, $\bnull$ denotes the all-zero word. We shall use the standard notation $\dist_L(\C)$ and $\dist^2_E(\C)$ to refer to the respective minimum distances (taken over all distinct codewords) of a code $\C\subseteq\Z_{2^h}^n$. The minimum squared Euclidean distance of a code essentially determines the performance of the code when employed for transmission over a white Gaussian noise channel at high signal-to-noise ratios.
\begin{theorem}
\label{thm:distance}
We have
\begin{align*}
\dist_L(\ERM(r,m,h))&=2^{m-r}\\
\dist^2_E(\ERM(r,m,h))&=2^{m-r+2}\sin^2\left(\frac{\pi}{2^h}\right).
\end{align*}
\end{theorem}
\vspace{1ex}
\begin{proof}
Since $\ERM(r,m,h)$ is linear, its minimum Lee distance is equal to the minimum Lee weight of the nonzero codewords. We shall first find a lower bound for the minimum Lee weight. It is then shown at the end of the proof that this bound is tight. Since $\ERM(r,m,1)$ is the classical Reed--Muller code, the theorem holds for $h=1$ (cf. \cite{MacWilliams1977}). This case serves as the anchor for the following induction.  Let $\bia=(a_0,a_1,\dots,a_{n-1})$ be a nonzero codeword in $\ERM(r,m,h)$. For $h>1$ let $b_i=a_i\pmod {2^{h-1}}$. Then $\bib=(b_0,b_1,\dots,b_{n-1})$ is a codeword in $\ERM(r,m,h-1)$. Since $a_i\in\{b_i,b_i+2^{h-1}\}$, it holds $\min\{a_i,2^h-a_i\}\ge\min\{b_i,2^{h-1}-b_i\}$, and therefore, $\wt_L(\bia)\ge\wt_L(\bib)\ge 2^{m-r}$, by induction on $h$.
\par
Now let us prove a lower bound on the minimum squared Euclidean distance. Again we have to find the minimum of $\wt^2_E(\bia)$ taken over all nonzero words $\bia\in\ERM(r,m,h)$. 
For any $u\in\Z_{2^h}$ we have
\begin{align*}
|\xi^u-1|^2&=4\sin^2\left(u\frac{\pi}{2^h}\right)\\
&=4\sin^2\left(\wt_L(u)\frac{\pi}{2^h}\right).
\end{align*}
For $1\le w\le 2^{h-1}$ it can be shown that
\beq
\sin^2\left(w\frac{\pi}{2^h}\right)\ge w\sin^2\left(\frac{\pi}{2^h}\right).
\eeq
Let $N_\bia(w)$ denote the number of entries in $\bia$ with Lee weight equal to $w$. Indeed
\begin{align}
\label{eqn:euclidean_weight}
\wt^2_E(\bia)&=4\sum_{w=1}^{2^{h-1}}N_\bia(w)\sin^2\left(w\frac{\pi}{2^h}\right)\nonumber\\
&\ge 4\sin^2\left(\frac{\pi}{2^h}\right)\sum_{w=1}^{2^{h-1}}wN_\bia(w)\\
&=4\sin^2\left(\frac{\pi}{2^h}\right)\wt_L(\bia).\nonumber
\label{eqn:de2}
\end{align}
It remains to exhibit a codeword in $\ERM(r,m,h)$, for which lower bounds are tight. Such a word is, for example, the word associated with the monomial $x_0\,x_1\cdots x_r$. This word has Lee weight $2^{m-r}$, and since it only contains zeros and ones, equality holds in (\ref{eqn:euclidean_weight}).
\end{proof}
\par
Next we relate $\ERM(r,m,h)$ to the codes $\RM_{2^h}(r,m)$ and $\ZRM_{2^h}(r,m)$ given in \cite{Davis1999}. These codes also generalize the binary Reed--Muller code to linear codes over $\Z_{2^h}$. We have
\beq
\RM_{2^h}(r,m)\subseteq \ERM(r,m,h),
\eeq
where the inclusion is proper if $h>1$ and $r<m$. Hence for $h>1$ and $r<m$ the code $\ERM(r,m,h)$ contains more codewords than $\RM_{2^h}(r,m)$, while both codes have minimum Lee distance equal to $2^{m-r}$.
For $h\ge 2$
\beq
\ZRM_{2^h}(r+1,m)\subseteq \ERM(r,m,h),
\eeq
which is a proper inclusion if $h>2$ and $r<m-1$. Hence for $h>2$ and $r<m-1$ the code $\ERM(r,m,h)$ contains more codewords than $\ZRM_{2^h}(r+1,m)$, while their minimum Lee distances are equal to $2^{m-r}$.


\subsection{OFDM Code Constructions}

We begin with defining a linear code over $\Z_{2^h}$.
\begin{definition}
\label{def:A}
For $0\le k<m$, $0\le r\le k+1$, and $h\ge 1$ we define the code $\A(k,r,m,h)$ to be the set of words corresponding to the set of polynomials
\begin{multline*}
\Bigg\{\sum_{i=0}^{m-k-1}x_\alpha g_i(x_{m-k},\dots,x_{m-1})+g(x_{m-k},\dots,x_{m-1})\,\bigg|\,\\
g_0,\dots,g_{m-k-1}\in\F(r-1,k,h),g\in\F(r,k,h)\Bigg\}.
\end{multline*}
\end{definition}
\vspace{1ex}
\par
Notice that $\A(0,1,m,h)$ is equal to the generalized first-order Reed--Muller code $\RM_{2^h}(1,m)$, described in \cite{Davis1999}. It follows from Definition~\ref{def:A} that $\A(k,r,m,h)$ is a linear code over $\Z_{2^h}$. Moreover
\beq
\A(k,r,m,h)\subseteq\ERM(r,m,h),
\eeq
and therefore, the minimum distances of $\A(k,r,m,h)$ can be lower-bounded with Theorem~\ref{thm:distance}. We remark that, similarly as in the proof of Theorem~\ref{thm:distance}, a particular word in $\A(k,r,m,h)$ can be identified showing that the lower bounds are in fact tight. 
The number of codewords in $\A(k,r,m,h)$ is equal to $2^s$, where
\beqn
\label{eqn:num_codewords_A}
s=(m-k)\cdot\log_2\big|\F(r-1,k,h)\big|+\log_2\big|\F(r,k,h)\big|,
\eeqn
which can be computed with (\ref{eqn:num_F}).
\par
As an example consider $\A(1,0,3,3)$. This code is a linear subcode of $\ERM(0,3,3)$ and has a generator matrix:
\beq
\begin{bmatrix}
1 & 1 & 1 & 1 & 1 & 1 & 1 & 1\\
0 & 2 & 0 & 2 & 0 & 2 & 0 & 2\\
0 & 0 & 2 & 2 & 0 & 0 & 2 & 2\\
0 & 0 & 0 & 0 & 2 & 2 & 2 & 2\\
0 & 0 & 0 & 0 & 0 & 4 & 0 & 4\\
0 & 0 & 0 & 0 & 0 & 0 & 4 & 4\\
\end{bmatrix}
\begin{matrix}
1\\2x_0\\2x_1\\2x_2\\4x_0x_2\phantom{.}\\4x_1x_2.
\end{matrix}
\eeq
Now let $\R(k,m,h)$ be the set of words associated with the following polynomials over $\Z_{2^h}$
\beq
\label{eqn:def-R}
2^{h-1}\sum_{\bid\in\{0,1\}^k}\sum_{i=0}^{m-k-2}x_{\pi_\bid(i)}x_{\pi_\bid(i+1)}\prod_{j=0}^{k-1}x_{m-k+j}^{d_j}(1-x_{m-k+j})^{(1-d_j)},
\eeq
where $\bid=(d_0,d_1,\dots,d_{k-1})$ and $\pi_{\bid}$ are $2^k$ permutations of $\{0,1,\dots,m-k-1\}$.
\begin{corollary}
\label{cor:cosets}
The corresponding polyphase words in the cosets of $\A(k,r,m,h)$ with coset representatives in $\R(k,m,h)$ have PMEPR at most $2^{k+1}$.
\end{corollary}
\begin{proof}
The corollary is a consequence of Theorem~\ref{thm:sets} and the following observations. By restricting any function corresponding to a word in $\A(k,r,m,h)$ in the variables $x_{m-k},\dots,x_{m-1}$, we obtain an affine function, and by restricting any function associated with a word in $\R(k,m,h)$ in the same variables, we obtain a quadratic polynomial, whose graph is a path of length $m-k$.
\end{proof}
\par
We are now in a position to construct a simple code.
\begin{construction}
\label{con:1}
Take a single coset of $\A(k,r,m,h)$ that contains a word in $\R(k,m,h)$. The polyphase versions of the words in this code have PMEPR at most $2^{k+1}$. The code has minimum Lee and squared Euclidean distance equal to $2^{m-r}$ and $2^{m-r+2}\sin^2\left(\frac{\pi}{2^h}\right)$, respectively, and the number of encoded bits per codeword is equal to $s=\log_2|\A(k,r,m,h)|$, which is given in  (\ref{eqn:num_codewords_A}).
\end{construction}
\par
In order to obtain a more elaborate code construction, we prove:
\begin{lemma}
For $m-k>1$ and $r>2-h$ the set $\R(k,m,h)$ contains 
\beqn
\label{eqn:eff_deg_R}
\left[\frac{(m-k)!}{2}\right]^{2^{\min\{r+h-3,k\}}}
\eeqn
words corresponding to a generalized Boolean function of effective degree at most $r$.
\end{lemma}
\begin{proof}
The set $\R(k,m,h)$ contains exactly $[(m-k)!/2]^{2^k}$ words, all having effective degree at most $k+3-h$. Hence the lemma is true for $r\ge k+3-h$. It is also clear that the expression in (\ref{eqn:def-R}) has algebraic degree at least $2$, so the effective degree is at least $3-h$. Now suppose that $3-h\le r<k+3-h$, and write $\ell=r+h-3$, where $0\le\ell<k$. By factoring out terms in the outer sum in (\ref{eqn:def-R}), it can be verified that, if 
\beq
\pi_{(d_0,\dots,d_{\ell-1},\,d_{\ell},\dots,d_{k-1})}=\pi_{(d_0,\dots,d_{\ell-1},\, 1-d_{\ell},\dots,1-d_{k-1})},
\eeq
then (\ref{eqn:def-R}) is independent of the variables $x_{m-k+\ell},\dots,x_{m-1}$ and, therefore, has effective degree at most $r=\ell+3-h$. This leaves the choice of $2^\ell=2^{r+h-3}$ permutations of the symbols $\{0,1,\dots,m-k-1\}$ that are distinct under reversal (e.g., all permutations satisfying $\pi(0)<\pi(m-k-1)$) to obtain distinct words in $\R(k,m,h)$ with effective degree at most $\ell+3-h$. This leads in total to the number given in (\ref{eqn:eff_deg_R}). 
\end{proof}
\begin{construction}
\label{con:2}
Suppose $m-k>1$. Let $2\le r\le k+2$ when $h=1$ and $1\le r\le k+1$ when $h>1$. Write $r'=\min\{r,k+1\}$. Let $2^{t}$ be the largest power of $2$ not exceeding (\ref{eqn:eff_deg_R}). Now take the union of $2^{t}$ distinct cosets of $\A(k,r',m,h)$, each containing a word in $\R(k,m,h)$ with effective degree at most $r$. The PMEPR of the corresponding polyphase words in this code is at most $2^{k+1}$, and one can encode $s+t$ bits, where $s=\log_2|\A(k,r',m,h)|$. Since the code is a subcode of $\ERM(r,m,h)$, its minimum Lee and squared Euclidean distance is at least $2^{m-r}$ and $2^{m-r+2}\sin^2\left(\frac{\pi}{2^h}\right)$, respectively. These are tight bounds if $r=r'$.
\end{construction}
\par
We remark that, when $k=0$, Construction~\ref{con:2} essentially restates the construction by Davis and Jedwab \cite{Davis1999}. A list of coding options having PMEPR at most 4 and at most 8 is compiled in Tables~\ref{tab:coding_options_4} and \ref{tab:coding_options_8}, respectively. 
The quantities $\dist_L$ and $\dist_E^2$ indicate lower bounds for the minimum Lee and the minimum squared Euclidean distance of the codes, respectively. The code with rate $R_1=s/2^m$ is obtained with Construction~\ref{con:1}, and the code with rate $R_2=(s+t)/2^m$ arises from Construction~\ref{con:2}. Notice that our definition of the code rate differs from the common one $\log_{2^h}|\C|/2^m$. The present definition has the advantage that it allows a fair comparison of codes over different alphabets on the basis of code rate and minimum squared Euclidean distance. 
\par
Finally, we wish to sketch how the proposed codes can be generally encoded and decoded. Encoding of the code $\A$ is straightforward by using a generator matrix for $\A$. Encoding of a union of cosets of $\A$ can be performed by using the information symbols partly to encode a word from $\A$ and partly to select a coset representative from a stored list. For decoding one needs to have an efficient algorithm to decode the linear code $\A$. This already provides a decoder for the codes from Construction~\ref{con:1}. Then codes from Construction~\ref{con:2} can be decoded by applying the supercode decoding method, as described in \cite{Conway1986} and \cite{Davis1999}. Such a concept involves subtracting all possible coset representatives from the received word in turn, and passing the resulting words to a decoder for the code $\A$. Among those decoder outputs the word that is closest to the received word determines the final decoding result.
\par
\setlength{\tabcolsep}{15pt}
\begin{table}[htb]
\centering
\caption{Coding Options with PMEPR at most 4}
\label{tab:coding_options_4}
\begin{tabular}{c|c|c|c|c|r@{.}l|r@{.}l|c|r@{.}l}
\hline
$m$ & $h$ & $r$ & $s$ & $t$ & \multicolumn{2}{c|}{$R_1$} & \multicolumn{2}{c|}{$R_2$} & $\dist_L$ & \multicolumn{2}{c}{$\dist^2_E$}\\ \hline\hline
4 & 1 & 2 & 8  & 1 & 0&50& 0&56 & 4 & 16&00\\
  &   & 3 & 8  & 3 & \multicolumn{2}{c|}{---} & 0&69 & 2 & 8&00\\
  & 2 & 1 & 13 & 1 & 0&81& 0&88 & 8 & 16&00\\
  &   & 2 & 16 & 3 & 1&00& 1&19 & 4 & 8&00\\
  & 3 & 1 & 21 & 3 & 1&31& 1&50 & 8 & 4&69\\
  &   & 2 & 24 & 3 & 1&50& 1&69 & 4 & 2&34\\\hline

5 & 1 & 2 & 10 & 3 & 0&31& 0&41 & 8 & 32&00\\
  &   & 3 & 10 & 7 & \multicolumn{2}{c|}{---} & 0&53 & 4 & 16&00\\
  & 2 & 1 & 16 & 3 & 0&50& 0&59 & 16 & 32&00\\
  &   & 2 & 20 & 7 & 0&63& 0&84 & 8 & 16&00\\
  & 3 & 1 & 26 & 7 & 0&81& 1&03 & 16 & 9&37\\
  &   & 2 & 30 & 7 & 0&94& 1&16 & 8 & 4&69\\\hline

6 & 1 & 2 & 12 & 5 & 0&19& 0&27 & 16 & 64&00\\
  &   & 3 & 12 & 11& \multicolumn{2}{c|}{---} & 0&36 & 8  & 32&00\\
  & 2 & 1 & 19 & 5 & 0&30& 0&38 & 32 & 64&00\\
  &   & 2 & 24 & 11& 0&38& 0&55 & 16 & 32&00\\ 
  & 3 & 1 & 31 & 11& 0&48& 0&66 & 32 & 18&75\\
  &   & 2 & 36 & 11& 0&56& 0&73 & 16 & 9&37\\ \hline 
\end{tabular}
\end{table}
\begin{table}[htb]
\centering
\caption{Coding Options with PMEPR at most 8}
\label{tab:coding_options_8}
\begin{tabular}{c|c|c|c|c|r@{.}l|r@{.}l|c|r@{.}l}
\hline
$m$ & $h$ & $r$ & $s$ & $t$ & \multicolumn{2}{c|}{$R_1$} & \multicolumn{2}{c|}{$R_2$} & $\dist_L$ & \multicolumn{2}{c}{$\dist^2_E$}\\ \hline\hline
5 & 1 & 2 & 13 & 1 & 0&41& 0&44 & 8 & 32&00\\
  &   & 3 & 16 & 3 & 0&50& 0&59 & 4 & 16&00\\
  &   & 4 & 16 & 6 & \multicolumn{2}{c|}{---} & 0&69 & 2 & 8&00\\
  & 2 & 1 & 19 & 1 & 0&59& 0&63 & 16& 32&00\\
  &   & 2 & 29 & 3 & 0&91& 1&00 & 8 & 16&00\\
  &   & 3 & 32 & 6 & 1&00& 1&19 & 4 & 8&00\\
  & 3 & 1 & 35 & 3 & 1&09& 1&19 & 16 & 9&37\\
  &   & 2 & 45 & 6 & 1&41& 1&59 & 8  & 4&69\\
  &   & 3 & 48 & 6 & 1&50& 1&69 & 4  & 2&34\\\hline

6 & 1 & 2 & 16 & 3 & 0&25& 0&30 & 16& 64&00\\
  &   & 3 & 20 & 7 & 0&31& 0&42 & 8 & 32&00\\
  &   & 4 & 20 & 14& \multicolumn{2}{c|}{---} & 0&53 & 4 & 16&00\\
  & 2 & 1 & 23 & 3 & 0&36& 0&41 & 32& 64&00\\
  &   & 2 & 36 & 7 & 0&56& 0&67 & 16& 32&00\\
  &   & 3 & 40 & 14& 0&63& 0&84 & 8 & 16&00\\
  & 3 & 1 & 43 & 7 & 0&67& 0&78 & 32& 18&75\\
  &   & 2 & 56 & 14& 0&88& 1&09 & 16& 9&37\\
  &   & 3 & 60 & 14& 0&94& 1&16 & 8 & 4&69\\\hline
\end{tabular}
\end{table}


\section{Discussion and Relations to Previous Constructions}
\label{sec:discussion}

It can be observed that QPSK (quaternary PSK) codes are always better than BPSK (binary PSK) codes, i.e., we can always construct a QPSK code with higher code rate and the same minimum Euclidean distance as a BPSK code. By moving to larger alphabets, the code rate can be increased further, but only at the cost of a smaller minimum Euclidean distance.
\par
It should be noted that Corollary~\ref{cor:cosets} and the arising code constructions do not exploit Theorem~\ref{thm:sets} in the most general way. The generalized Boolean functions corresponding to the words in the cosets identified in Corollary~\ref{cor:cosets} are characterized by the property that by restricting the functions in the variables $x_{m-k},\dots,x_{m-1}$, we obtain quadratic functions whose graphs are paths in the vertices $0,\dots,m-k-1$. In order to increase the size of the codes in Constructions~\ref{con:1} and \ref{con:2}, we can, according to Theorem~\ref{thm:sets}, apply any permutation to the $m$ variables in the functions corresponding to the codewords (instead of only to a fixed set of $m-k$ variables). This, however, has the unwanted effect that some codewords are generated more than once. Such an approach, coupled with rather complicated techniques to remove multiple codewords, has been used in \cite{Paterson2000a}, where the functions are constrained to have quadratic degree. Our approach has the advantage that these difficulties are avoided. Moreover, compared to the concept in \cite{Paterson2000a}, it allows us to construct our codes as unions of relatively few cosets of a relatively large linear code, which presumably simplifies the decoding process. The penalty of this simplification is a loss of at most $\log_2{m\choose k}$ encodable information bits (since, instead of ${m\choose k}$ possible index sets, we choose just one set of $m-k$ indices that form the vertices of the paths in the graphs of the restricted functions). This loss is moderate for typical choices of $m$ and $k$.
\par
Finally we wish to compare the codes arising from Construction~\ref{con:2} with those in \cite{Davis1999} and \cite{Paterson2000a}. The codes in the latter references are contained in $\RM_{2^h}(2,m)$ or $\ZRM_{2^h}(2,m)$, which ensures a minimum Lee distance of at least $2^{m-2}$ or $2^{m-1}$, respectively. So we compare these codes with codes arising from Construction~\ref{con:2} having the same lower bound on the minimum Lee distance, i.e., we let $r\in\{1,2\}$. Also we let $k=1$, which covers the majority of the codes in \cite{Davis1999} and \cite{Paterson2000a} having PMEPR greater than $2$.
\par
For $h=1$, $r=2$, and $m\ge3$ the code from Construction~\ref{con:2} can be used to encode 
\beq
\lfloor\log_2(m-1)!\rfloor+2m-1
\eeq
bits. This yields 13 and 17 bits for $m=5$ and $m=6$, respectively. These values should be compared with 11 and 17 bits in \cite[Table I]{Paterson2000a}, which suggests that for $h=1$ and for small $m$ our construction is slightly stronger than that in \cite{Paterson2000a}. However for $h=1$ and $m\ge 8$ it was stated in \cite{Paterson2000a} that the number of encoded bits of the codes in \cite{Paterson2000a} is equal to $\lfloor\log_2m!\rfloor+2m-2$. Hence for large $m$ the binary code from \cite{Paterson2000a} allows to encode either $\lfloor\log_2 m\rfloor$ or $\lfloor\log_2 m\rfloor-1$ bits more than a comparable code arising from Construction~\ref{con:2}. We arrive at a similar conclusion for $h=2$ and $r=1$. For $m\ge3$ we can encode 
\beq
\lfloor\log_2(m-1)!\rfloor+3m
\eeq
bits, which is, compared to a code in \cite{Paterson2000a} with the same minimum distance, slightly larger for small $m$ and is either $\lfloor\log_2 m\rfloor$ or $\lfloor\log_2 m\rfloor-1$ bits less for $m\ge 8$. 
\par
For $h\ge 2$, $r=2$, and $m\ge3$ Construction~\ref{con:2} yields a code, which can be used to encode 
\beq
\lfloor2\cdot\log_2(m-1)!\rfloor+2hm-2
\eeq
bits. When $m\ge 4$, the number of encoded bits for a comparable code in \cite{Paterson2000a} is equal to $\lfloor\log_2m!\rfloor+2hm-2$.
This is $\lfloor\log_2m!-2\cdot\log_2m\rfloor$ or $\lfloor\log_2m!-2\cdot\log_2m\rfloor+1$ bits less than the code from Construction~\ref{con:2}. Similar results can be established for $h>2$ and $r=1$.
\par
In summary, except for large $m$ in the cases $(h,r)=(1,2)$ and $(h,r)=(2,1)$, the codes from Construction~\ref{con:2} outperform coding schemes proposed proposed in \cite{Paterson2000a}.
\par
Based on exhaustive computational search, \cite{Davis1999} reports codes that outperform the codes given in the first and the third row of Table~\ref{tab:coding_options_4} by one encoded information bit and the codes in the seventh and ninth row of Table~\ref{tab:coding_options_4} by two encoded information bits. These observations can be partly explained using a variety of individual theorems from \cite{Schmidt2006c}, \cite{Paterson2000a}, \cite{Stinchcombe2000}, \cite{Chen2006} and show that stronger constructions are possible in some situations. However the description of such codes (and therefore encoding and decoding) tends to be unwieldy.




\begin{thebibliography}{10}
\providecommand{\url}[1]{#1}
\csname url@rmstyle\endcsname
\providecommand{\newblock}{\relax}
\providecommand{\bibinfo}[2]{#2}
\providecommand\BIBentrySTDinterwordspacing{\spaceskip=0pt\relax}
\providecommand\BIBentryALTinterwordstretchfactor{4}
\providecommand\BIBentryALTinterwordspacing{\spaceskip=\fontdimen2\font plus
\BIBentryALTinterwordstretchfactor\fontdimen3\font minus
  \fontdimen4\font\relax}
\providecommand\BIBforeignlanguage[2]{{%
\expandafter\ifx\csname l@#1\endcsname\relax
\typeout{** WARNING: IEEEtran.bst: No hyphenation pattern has been}%
\typeout{** loaded for the language `#1'. Using the pattern for}%
\typeout{** the default language instead.}%
\else
\language=\csname l@#1\endcsname
\fi
#2}}

\bibitem{Calderbank1995a}
A.~R. Calderbank and N.~J.~A. Sloane, ``Modular and $p$-adic cyclic codes,''
  \emph{Designs, Codes and Cryptography}, vol.~6, pp. 21--35, 1995.

\bibitem{Chen2006}
C.-Y. Chen, C.-H. Wang, and C.-C. Chao, ``Complementary sets and {Reed--Muller}
  codes for peak-to-average power ratio reduction in {OFDM},'' \emph{Proc. of
  16th AAECC Symp. (Lecture Notes in Computer Science)}, vol. 3857, pp.
  317--327, 2006.

\bibitem{Conway1986}
J.~H. Conway and N.~J.~A. Sloane, ``Soft decoding techniques for codes and
  lattices, including the {Golay} code and the {Leech} lattice,'' \emph{IEEE
  Trans. Inf. Theory}, vol.~32, no.~1, pp. 41--50, Jan. 1986.

\bibitem{Davis1999}
J.~A. Davis and J.~Jedwab, ``Peak-to-mean power control in {OFDM}, {Golay}
  complementary sequences, and {Reed--Muller} codes,'' \emph{IEEE Trans. Inf.
  Theory}, vol.~45, no.~7, pp. 2397--2417, Nov. 1999.

\bibitem{Golay1961}
M.~J.~E. Golay, ``Complementary series,'' \emph{IRE Trans. Inf. Theory},
  vol.~7, no.~2, pp. 82--87, Apr. 1961.

\bibitem{Hammons1994}
A.~R. Hammons, P.~V. Kumar, A.~R. Calderbank, N.~J.~A. Sloane, and P.~Sol\'{e},
  ``The {${\mathbb Z}_4$}-linearity of {Kerdock, Preparata, Goethals} and
  related codes,'' \emph{IEEE Trans. Inf. Theory}, vol.~40, no.~2, pp.
  301--319, Mar. 1994.

\bibitem{Jones1996}
A.~E. Jones and T.~A. Wilkinson, ``Combined coding for error control and
  increased robustness to system nonlinearities in {OFDM},'' \emph{Proc. of
  IEEE 46th Vehicular Technology Conf. (VTC), Atlanta, GA}, pp. 904--908, Apr.
  1996.

\bibitem{Jones1994}
A.~E. Jones, T.~A. Wilkinson, and S.~K. Barton, ``Block coding scheme for
  reduction of peak to mean envelope power ratio of multicarrier transmission
  schemes,'' \emph{IEE Electron. Lett.}, vol.~30, no.~25, pp. 2098--2099, Dec.
  1994.

\bibitem{MacWilliams1977}
F.~J. MacWilliams and N.~J.~A. Sloane, \emph{The Theory of Error-Correcting
  Codes}.\hskip 1em plus 0.5em minus 0.4em\relax North Holland Mathematical
  Library, 1977.

\bibitem{Ochiai1997}
H.~Ochiai and H.~Imai, ``Block coding scheme based on complementary sequences
  for multicarrier signals,'' \emph{IEICE Transactions on Fundamentals}, vol.
  E80-A, pp. 2136--2146, Nov. 1997.

\bibitem{Parker2003}
\BIBentryALTinterwordspacing
M.~G. Parker and C.~Tellambura, ``A construction for binary sequence sets with
  low peak-to-average power ratio,'' \emph{Report No. 242, Department of
  Informatics, University of Bergen, Norway}, Feb. 2003. [Online]. Available:
  \url{http://www.ii.uib.no/~matthew/ConstructReport.pdf}
\BIBentrySTDinterwordspacing

\bibitem{Paterson2000a}
K.~G. Paterson, ``Generalized {Reed--Muller} codes and power control in {OFDM}
  modulation,'' \emph{IEEE Trans. Inf. Theory}, vol.~46, no.~1, pp. 104--120,
  Jan. 2000.

\bibitem{Paterson2000}
K.~G. Paterson and V.~Tarokh, ``On the existence and construction of good codes
  with low peak-to-average power ratios,'' \emph{IEEE Trans. Inf. Theory},
  vol.~46, no.~6, pp. 1974--1987, Sep. 2000.

\bibitem{Popovic1991}
B.~M. Popovi\'{c}, ``Synthesis of power efficient multitone signals with flat
  amplitude spectrum,'' \emph{IEEE Trans. Commun.}, vol.~39, no.~7, pp.
  1031--1033, Jul. 1991.

\bibitem{Schmidt2006c}
K.-U. Schmidt, ``On cosets of the generalized first-order {Reed--Muller} code
  with low {PMEPR},'' \emph{IEEE Trans. Inf. Theory}, vol.~52, no.~7, Jul.
  2006.

\bibitem{Stinchcombe2000}
\BIBentryALTinterwordspacing
T.~E. Stinchcombe, ``Aperiodic autocorrelations of length {$2^m$} sequences,
  complementarity, and power control for {OFDM},'' Ph.D. dissertation,
  University of London, Apr. 2000. [Online]. Available:
  \url{http://www.isg.rhul.ac.uk/alumni/thesis/stinchcombe_t.pdf}
\BIBentrySTDinterwordspacing

\bibitem{Nee1996}
R.~D.~J. van Nee, ``{OFDM} codes for peak-to-average power reduction and error
  correction,'' \emph{Proc. of IEEE Global Telecommunications Conference
  (GLOBECOM), London, U.K.}, pp. 740--744, Nov. 1996.

\bibitem{Wilkinson1995}
T.~A. Wilkinson and A.~E. Jones, ``Minimisation of the peak to mean envelope
  power ratio of multicarrier transmission schemes by block coding,''
  \emph{Proc. of IEEE Vehicular Technology Conference (VTC), Chicago, IL}, pp.
  825--829, Jul. 1995.

\end{thebibliography}
\end{document}